\documentclass[twocolumn,english,showpacs,aps,superscriptaddress,prl]{revtex4}
\usepackage[T1]{fontenc}
\usepackage[latin1]{inputenc}
\usepackage{amsmath}
\usepackage{graphicx}
\usepackage{amssymb}

\makeatletter
\usepackage{hyperref}
\hypersetup{
breaklinks=true,
colorlinks=true,
citecolor=blue,
linkcolor=blue,
filecolor=blue,
urlcolor=blue
}

\usepackage{babel}
\makeatother
\begin{document}

\title{Theory of smeared quantum phase transitions}

\author{Jos\'{e} A. Hoyos}

\affiliation{Department of Physics, Duke University, Durham, NC 27708, USA}

\affiliation{Department of Physics, Missouri University of Science and Technology,
Rolla, MO 65409, USA}

\author{Thomas Vojta}

\affiliation{Department of Physics, Missouri University of Science and Technology,
Rolla, MO 65409, USA}

\begin{abstract}
We present an analytical strong-disorder renormalization group theory
of the quantum phase transition in the dissipative random transverse-field
Ising chain. For Ohmic dissipation, we solve the renormalization flow
equations analytically, yielding asymptotically exact results for
the low-temperature properties of the system. We find that the interplay
between quantum fluctuations and Ohmic dissipation destroys the quantum
critical point by smearing. We also determine the phase diagram and
the behavior of observables in the vicinity of the smeared quantum
phase transition. 
\end{abstract}


\maketitle
One of the most basic questions concerning phase transitions in random
systems is whether or not a sharp transition survives in the presence
of quenched disorder. Initially, it was suspected that disorder destroys
any critical point because different spatial regions order at different
temperatures. However, it was soon realized that classical continuous
phase transitions generically remain sharp in the presence of weak
disorder because finite spatial regions cannot undergo a true phase
transition (see Ref.~\cite{grinstein} and references therein). 

Nonetheless, rare strongly coupled spatial regions play an important
role. They can be locally in the ordered phase even if the bulk system
is in the disordered phase. The slow fluctuations of these regions
give rise to a singular free energy in a whole temperature region
around the transition (called the Griffiths phase) \cite{griffiths-prl69,mccoy-prl69}.
In generic classical systems, this is a weak effect, because the Griffiths
singularity is only an essential one. In contrast, rare regions can
play a more important role at zero-temperature quantum phase transitions
where order-parameter fluctuations in space and (imaginary) time need
to be considered. Quenched disorder is perfectly correlated in time
direction, and this enhances the Griffiths singularities. In the prototypical
random transverse-field Ising systems, the singularities take power-law
forms, implying, e.g., a divergent susceptibility in the Griffiths
phase \cite{thill-huse-physa95,young-rieger-prb96,guo-bhatt-huse-prb96}.
The transition itself is governed by an exotic infinite-randomness
critical point \cite{fisher92,fisher95}, but remains sharp. 

Recently, it was noted that dissipation can further enhance rare region
effects at quantum phase transitions with Ising order parameter symmetry.
Each locally ordered region acts as two-level system. When coupled
to an Ohmic dissipative bath, it can undergo the localization transition
of the spin-boson problem \cite{leggett-etal-rmp87}. Thus, each region
can order independently of the bulk system, destroying the sharp phase
transition by smearing \cite{vojta-prl03}. In view of this observation,
it would be highly desirable to treat the nonperturbative physics
of these dissipative rare regions within the framework of the renormalization
group (RG) commonly used to describe phase transitions. Such a theory
would not only unveil the ultimate fate of the critical point, it
would also predict \emph{quantitatively} the behavior of many observables
near the transition. 

An important step towards this goal was taken by Schehr and Rieger
\cite{schehr-rieger06,scherh-rieger-jsm08} who studied the dissipative
random transverse-field Ising chain by a numerical strong-disorder
RG. They confirmed the smeared transition scenario and focused on
the infinite-randomness {}``pseudo''-critical point arising at intermediate
energy scales where dissipative effects are less important. 

\begin{figure}
\begin{center}\includegraphics[%
  clip,
  width=0.9\columnwidth,
  keepaspectratio]{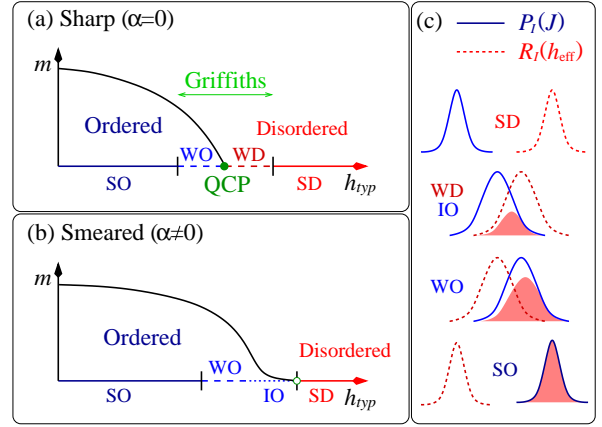}\end{center}

\caption{(Color online) Zero temperature phase diagram and magnetization of
the dissipative random transverse-field Ising chain as a function
of the typical transverse field $h_{typ}$. SO and SD denote the strongly
ordered and disordered conventional phases; WO and WD are the weakly
ordered and disordered quantum Griffiths phases. (a) No dissipation:
sharp QCP. (b) Ohmic dissipation: smeared transition with the inhomogeneously
ordered (IO) phase replacing the WD Griffiths phase. (c) Distributions
of the bonds $J$ and fields ${h_{\textrm{eff}}}$ in the various
phases. The shaded area quantifies the fraction $w$ of $J$'s bigger
than ${h_{\textrm{eff}}}$'s {[}see Eq.~(\ref{eq:w-definition}){]}.
\label{cap:phase-diagram}}
\end{figure}

In this Letter, we develop a comprehensive strong-disorder RG for
the dissipative random transverse-field Ising chain. We derive RG
flow equations for the distributions of the fields, bonds and magnetic
moments and solve them analytically, providing asymptotically exact
low-energy results. We prove that the quantum critical point (QCP)
is destroyed by Ohmic dissipation. Instead, a smeared quantum phase
transition separates a conventional paramagnet from an inhomogeneously
ordered ferromagnet (Fig.~\ref{cap:phase-diagram}). In the remainder
of the Letter, we sketch the derivation of our theory, compute important
observables, and discuss the relevance of our results. Extensive details
will be given in a longer paper. 

Our starting point is the dissipative random transverse-field Ising
chain defined by the Hamiltonian \begin{eqnarray}
H= & - & \sum_{i}J_{i}\sigma_{i}^{z}\sigma_{i+1}^{z}-\sum_{i}h_{i}\sigma_{i}^{x}\nonumber \\
 & + & \sum_{i,n}\sigma_{i}^{z}\lambda_{i,n}(a_{i,n}^{\dagger}+a_{i,n}^{\phantom{\dagger}})+\sum_{i,n}\nu_{i,n}a_{i,n}^{\dagger}a_{i,n}^{\phantom{\dagger}}\label{eq:H}\end{eqnarray}
 where $\sigma_{i}^{x,z}$ are Pauli matrices. The bonds $J_{i}$
and fields $h_{i}$ are independent random variables; $a_{i,n}^{\dagger}$
($a_{i,n}$) are the creation (annihilation) operators of the $n$-th
oscillator coupled to spin $\sigma_{i}$ via $\lambda_{i,n}$, and
$\nu_{i,n}$ is its frequency. Initially, all baths have the same
Ohmic spectral function ${\mathcal{E}}(\omega)=\pi\sum_{n}\lambda_{i,n}^{2}\delta(\omega-\nu_{i,n})=2\pi\alpha\omega e^{-\omega/\omega_{c}}$,
with $\alpha$ the dimensionless dissipation strength and $\omega_{c}$
the (bare) cutoff energy. (The cutoff will change under the RG and
the dissipation strength will become site-dependent.) 

To characterize the low-energy behavior of the system (\ref{eq:H}),
we now develop a strong-disorder RG \cite{MDH-PRL,igloi-review}.
The idea of this method is to successively integrate out local high-energy
modes. In our case, the competing energies are the transverse fields,
bonds, and oscillator frequencies. Each RG step proceeds as follows:
We first find the largest energy in the system $\Omega=\max(h_{i},J_{i},\omega_{c}/p)$
where $p\gg1$ is an arbitrary constant %
\cite{foot1}. We then lower the energy scale from $\Omega$ to $\Omega-{\textrm{d}}\Omega$
by (i) integrating out all oscillators (at all sites $i$) with frequencies
between $p(\Omega-{\textrm{d}}\Omega)$ and $p\Omega$ and (ii) decimating
all transverse fields and bonds between $(\Omega-{\textrm{d}}\Omega)$
and $\Omega$. 

For $p\gg1$, the oscillators can be treated using adiabatic renormalization
\cite{leggett-etal-rmp87}. As a result, the transverse fields renormalize
according to \begin{equation}
\tilde{h}_{i}=h_{i}\exp\left(-\alpha_{i}\int_{p(\Omega-{\textrm{d}}\Omega)}^{p\Omega}\frac{{\textrm{d}}\omega}{\omega}\right)=h_{i}\left(1-\alpha_{i}\frac{{\textrm{d}}\Omega}{\Omega}\right)\label{eq:h-tilde-bath}\end{equation}
 while the bonds remain unchanged. Here $\alpha_{i}$ is the renormalized
dissipation strength at site $i$. 

To decimate a strong bond $J_{i}=\Omega$, we assume the spins $\mathbf{\sigma}_{i}$
and $\mathbf{\sigma}_{i+1}$ to be locked together as an effective
spin cluster $\tilde{\mathbf{\sigma}}$ with moment $\tilde{\mu}$
and renormalized transverse field $\tilde{h}$ obtained in second
order perturbation theory, \begin{eqnarray}
\tilde{\mu} & = & \mu_{i}+\mu_{i+1}~,\label{eq:mu-tilde}\\
\tilde{h} & = & h_{i}h_{i+1}/J_{i}~.\label{eq:h-tilde-J}\end{eqnarray}
 $\tilde{\mathbf{\sigma}}$ couples to a renormalized bath of dissipation
strength \begin{equation}
\tilde{\alpha}=\alpha_{i}+\alpha_{i+1}=\alpha(\mu_{i}+\mu_{i+1})=\alpha\tilde{\mu}~.\label{eq:alfa-tilde}\end{equation}

For a strong field, $h_{i}=\Omega$, the corresponding spin $\mathbf{\sigma}_{i}$
is delocalized in $\sigma^{z}$ basis and thus eliminated, creating
a new bond between sites $i-1$ and $i+1$, \begin{equation}
\tilde{J}=J_{i-1}J_{i}/h_{i}~.\label{eq:J-tilde}\end{equation}
 Note that for spins about to be decimated, $h_{i}=\Omega$ is the
fully renormalized tunnel splitting $h_{i}=h_{i0}(p\, h_{i0}/\omega_{c0})^{\alpha\mu_{i}/(1-\alpha\mu_{i})}$
where $h_{i0}$ and $\omega_{c0}$ are the field and bath cutoff of
the $i-$th cluster when it was formed at the higher energy $\omega_{c0}/p$. 

The recursion relations (\ref{eq:mu-tilde}), (\ref{eq:h-tilde-J}) and
(\ref{eq:J-tilde}) are identical to the dissipationless case \cite{fisher92},
the baths enter only via (\ref{eq:h-tilde-bath}) together with the
renormalization of the dissipation strengths (\ref{eq:alfa-tilde}).
Our RG procedure is related to the one implemented numerically by
Schehr and Rieger \cite{schehr-rieger06}. However, treating the oscillator
modes on equal footing with the other degrees of freedom (by reducing
the bath cutoff globally in each step) allows us to solve the problem
analytically. 

The complete RG step consisting of recursion relations (\ref{eq:h-tilde-bath})--(\ref{eq:J-tilde})
is now iterated with the energy scale $\Omega$ being decreased. At
each stage, the remaining bonds $J$ and fields $h$ are independent,
but the fields and magnetic moments are correlated. Using logarithmic
variables $\Gamma=\ln(\Omega_{I}/\Omega)$ {[}where $\Omega_{I}$
is the initial (bare) value of $\Omega${]}, $\zeta=\ln(\Omega/J)$
and $\beta=\ln(\Omega/h)$, we can thus derive RG flow equations for
the bond distribution ${\mathcal{P}}(\zeta)$ and the joint distribution
of fields and moments ${\mathcal{R}}(\beta,\mu)$. They read \begin{eqnarray}
\frac{\partial{\cal P}}{\partial\Gamma} & = & \frac{\partial{\cal P}}{\partial\zeta}+\left(1-\alpha\overline{\mu}_{0}\right){\cal R}_{\beta}\left(0\right)\left({\cal P}\stackrel{\zeta}{\otimes}{\cal P}\right)\nonumber \\
 &  & +\left[{\cal P}\left(0\right)-\left(1-\alpha\overline{\mu}_{0}\right){\cal R}_{\beta}\left(0\right)\right]{\cal P}~,\label{eq:flow-P}\\
\frac{\partial{\cal R}}{\partial\Gamma} & = & \left(1-\alpha\mu\right)\frac{\partial{\cal R}}{\partial\beta}+{\cal P}\left(0\right)\left({\cal R}\stackrel{\beta,\mu}{\otimes}{\cal R}\right)\nonumber \\
 &  & -\left[{\cal P}\left(0\right)-\left(1-\alpha\overline{\mu}_{0}\right){\cal R}_{\beta}\left(0\right)\right]{\cal R}~,\label{eq:flow-R}\end{eqnarray}
where ${\mathcal{R}}_{\beta}(\beta)=\int_{0}^{\infty}{\mathcal{R}}(\beta,\mu){\textrm{d}}\mu$
is the distribution of the fields and $\overline{\mu}_{0}$ is the
average moment of clusters about to be decimated (defined by $\overline{\mu}_{0}{\mathcal{R}}_{\beta}(0)=\int_{0}^{\infty}\mu{\mathcal{R}}(0,\mu){\textrm{d}}\mu$).
The symbol ${\mathcal{P}}\stackrel{\zeta}{\otimes}{\mathcal{P}}=\int_{0}^{\zeta}{\mathcal{P}}(\zeta^{\prime}){\mathcal{P}}(\zeta-\zeta^{\prime}){\textrm{d}}\zeta^{\prime}$
denotes the convolution. The first term on the r.h.s. of (\ref{eq:flow-P})
and (\ref{eq:flow-R}) is due to the rescaling of $\zeta$ and $\beta$
with $\Gamma$ and the renormalization (\ref{eq:h-tilde-bath}) of
$h$ by the baths. The second term implements the recursion relations
(\ref{eq:mu-tilde}), (\ref{eq:h-tilde-J}) and (\ref{eq:J-tilde}) for the
moments, fields and bonds. The last term ensures the normalization
of $\mathcal{P}$ and $\mathcal{R}$. As expected, for $\alpha=0$,
(\ref{eq:flow-P}) and (\ref{eq:flow-R}) become identical to the
dissipationless case \cite{fisher92,fisher95}. 

Important insight can already be obtained from the structure of the
flow equations. The probability of decimating a field, $(1-\alpha\overline{\mu}_{0}){\mathcal{R}}_{\beta}(0)$,
decreases with increasing dissipation strength and cluster size. Clusters
with moment $\mu>1/\alpha$ are not decimated. Thus, in the presence
of dissipation, the flow equations always contain a \emph{finite}
length scale above which the cluster dynamics freezes. 

We now search for stationary solutions of the flow equations (\ref{eq:flow-P})
and (\ref{eq:flow-R}) that describe stable phases or critical points.
There are two trivial cases: If all bonds are larger than all fields,
only bonds are decimated, building larger and larger clusters. This
is the conventional strongly ordered (SO) ferromagnetic phase. If
only fields are decimated, we are in the conventional strongly disordered
(SD) paramagnetic phase. 

In the more interesting case of overlapping field and bond distributions,
we look for solutions invariant under a general rescaling $\eta=\zeta/f_{\zeta}(\Gamma)$,
$\theta=\beta/f_{\beta}(\Gamma)$ and $\nu=\mu/f_{\mu}(\Gamma)$. 

Without dissipation, $\alpha=0$, there are three types of well-behaved
solutions \cite{fisher95}: a line of fixed points (parameterized
by ${\mathcal{R}}_{0}$) with $f_{\beta}=1$, $f_{\zeta}=\exp({\mathcal{R}}_{0}\Gamma)$
and the average moment increasing as $\Gamma$. It corresponds to
the weakly disordered (WD) Griffiths phase. There is another line
of fixed points with $f_{\zeta}=1$ and $f_{\beta}=f_{\mu}=\exp({\mathcal{P}}_{0}\Gamma)$
(parameterized by ${\mathcal{P}}_{0}$) which corresponds to the weakly
ordered (WO) Griffiths phase; and, separating these two phases, an
infinite-randomness QCP with $f_{\zeta}=f_{\beta}=\Gamma$ and $f_{\mu}=\Gamma^{\phi}$,
with $2\phi=1+\sqrt{5}$. 

In the presence of dissipation, $\alpha\neq0$, the scenario changes
dramatically. For overlapping bond and field distributions, we found
only \emph{one} line of well-behaved fixed points (parameterized by
${\mathcal{P}}_{0}>0$) corresponding to the ordered phase %
\cite{foot2}. Here, $f_{\zeta}=1$, $f_{\mu}=\exp({\mathcal{P}}_{0}\Gamma)$,
$f_{\beta}=\Gamma\exp({\mathcal{P}}_{0}\Gamma)$. The fields become
much smaller than the bonds, justifying the perturbative treatment
of the RG step. The fixed-point distributions are \begin{subequations}
\begin{eqnarray}
{\mathcal{P}}^{*}(\zeta) &=& {\mathcal{P}}_{0}e^{-{\mathcal{P}}_{0}\zeta}~,\label{eq:fixed-point-p}\\
{\mathcal{R}}^{*}(\theta,\nu) &=& {\mathcal{R}}_{0} 
e^{-{\mathcal{R}}_{0}\nu} \delta(\theta - \alpha \nu )~,\label{eq:fixed-point-r}
\end{eqnarray}
\label{eqs:fixed-point}
\end{subequations} i.e., fields and moments are perfectly correlated. Here, ${\mathcal{R}}_{0}$
is a nonuniversal constant. This fixed point is similar to the WO
Griffiths phase for $\alpha=0$, but $f_{\beta}/f_{\mu}\rightarrow\infty$
as $\Gamma\rightarrow\infty$. Transforming the field distribution
(\ref{eq:fixed-point-r}) back to the original transverse fields $h$
gives power-law behavior $\sim h^{{\mathcal{R}}_{0}/(\alpha f_{\beta})-1}$.
We could not analytically solve for the nonuniversal constants ${\mathcal{P}}_{0}$
and ${\mathcal{R}}_{0}$ in terms of the bare distributions and $\alpha$.
Their numerical values will be given elsewhere. 

We emphasize that we have shown that there is no fixed point solution
with $f_{\zeta}/f_{\beta}\rightarrow{\textrm{const}}$ as $\Gamma\rightarrow\infty$
in the presence of dissipation, implying that there is no QCP where
fields and bonds compete at \emph{all} energy scales. This important
result proves that Ohmic dissipation destroys Fisher's \cite{fisher92,fisher95}
infinite-randomness critical point. Physically, it is due to the fact
that \emph{finite} spin clusters (of size $\sim1/\alpha$) can develop
true magnetic order. 

The complete low-energy thermodynamics can be obtained from the RG
fixed point solutions. To characterize the phase diagram (Fig.~\ref{cap:phase-diagram})
in terms of the bare variables we introduce the probability \begin{equation}
w=\int_{0}^{\infty}{\textrm{d}}JP_{I}(J)\int_{0}^{J}{\textrm{d}}h_{\textrm{eff}}{R}_{I}(h_{\textrm{eff}})~,\label{eq:w-definition}\end{equation}
 of a bare bond $J$ being greater than an effective field (a bare
field, fully renormalized by the baths) $h_{\textrm{eff}}=h(ph/\omega_{c})^{\alpha/(1-\alpha)}$.
$P_{I}(J)$ and $R_{I}(h_{\textrm{eff}})$ are the bare initial distributions
of these variables {[}see Fig.~\hyperref[cap:phase-diagram]{\ref{cap:phase-diagram}(c)}{]}. 

For $w=0$ and $w=1$, these distributions do not overlap. The system
is in one of the conventional phases (SD or SO) without Griffiths
singularities where disorder is RG irrelevant. For $0<w\ll1$, arbitrarily
large rare clusters can form under renormalization. Without dissipation,
$\alpha=0$, these clusters have small but nonzero effective fields.
They thus slowly fluctuate, and the system is in the WD Griffiths
phase {[}see Fig.~\hyperref[cap:phase-diagram]{\ref{cap:phase-diagram}(a)}{]}.
In the presence of dissipation, $\alpha\neq0$, clusters with moment
$\mu>1/\alpha$ have zero effective field. They freeze and order independently
from the bulk. The sharp transition is thus destroyed by smearing,
and the WD Griffiths phase is replaced by an inhomogeneously ordered
(IO) ferromagnetic phase {[}see Fig.~\hyperref[cap:phase-diagram]{\ref{cap:phase-diagram}(b)}{]}.
Finally, with $w$ approaching 1, the system develops bulk magnetic
order but rare fluctuating clusters still exist, i.e., we are in the
WO Griffiths phase. In the presence of dissipation, the IO and WO
phases are separated by a crossover rather than a QCP. The asymptotic
low-energy properties of both phases are described by the solution
(\ref{eqs:fixed-point}) with ${\mathcal{P}}_{0}$ monotonically decreasing
with $w$. 

We now turn our attention to observables near the smeared phase transition,
focusing on the IO ferromagnetic phase which is the novel feature
of our system. The magnetization is dominated by the large frozen
droplets which arise in rare regions where the bonds are greater than
the local fields. Because they are static, any weak coupling mediated
by the bulk is sufficient to align them. Hence, the magnetization
is proportional to the volume of the rare frozen droplets, which for
$\alpha$ and $w\ll1$, is \begin{equation}
m\sim w^{1/\alpha}~.\label{eq:mag}\end{equation}

The low-temperature magnetic susceptibility can be computed by running
the RG to energy scale $\Omega=T$ and assuming the remaining spin
clusters to be free. For asymptotically low energies, the RG flow
is dictated by the fixed-point solution (\ref{eqs:fixed-point}),
leading to \begin{equation}
\chi\sim T^{-1-1/z}~,\label{eq:low-chi}\end{equation}
 with $z=1/{\mathcal{P}}_{0}$. Note, however, that at higher energies,
the flow is dominated by strong fields and the susceptibility therefore
behaves as in the weakly disordered undamped Griffiths phase \cite{fisher95}:
\begin{equation}
\chi\sim\delta^{4-2\phi}\left[\ln\left(1/T\right)\right]^{2}T^{-1+1/z^{\prime}}~,\label{eq:high-chi}\end{equation}
 with $z^{\prime}\approx1/(2\delta)$, and $\delta\approx\left\langle \right.\!\!\ln{h_{\textrm{eff}}}\!\!\left.\right\rangle -\left\langle \ln J\right\rangle $.
The crossover energy $\Omega_{c}$ separating the two regimes can
be estimated as the energy in which the high-energy mean moment cluster,
$\overline{\mu}\sim\Gamma\delta^{1-\phi}$, reaches the critical size
$1/\alpha$. Hence, $\alpha\ln(\Omega_{I}/\Omega_{c})\sim\delta^{\phi-1}$.
Below $\Omega_{c}$, the mean magnetic moment increases much more
rapidly, $\overline{\mu}\sim\exp\left({\mathcal{P}}_{0}\Gamma\right)$. 

In summary, we have developed an asymptotically exact strong-disorder
RG theory for the dissipative random transverse-field Ising chain.
We have solved the resulting flow equations analytically and proven
that the QCP is destroyed by smearing. The smearing is the result
of the \emph{interplay} between disorder and dissipation. Dissipation
alone leads to a conventional critical point \cite{werner-etal-prl-05},
while disorder alone leads to an exotic infinite-randomness critical
point \cite{fisher92,fisher95}, but the transition remains sharp.
In the remaining paragraphs, we put our results in broader perspective,
and we discuss further implications. 

We first consider the dissipative random transverse-field Ising model
in higher dimensions. The recursion relations (\ref{eq:h-tilde-bath})--(\ref{eq:alfa-tilde})
are the same as in one dimension while decimating a field now generates
couplings between all nearest neighbor sites, changing the topology
of the lattice. An analytical solution of the RG flow equations thus
appears impossible. However, the dissipation terms are local and take
the same form as in one dimension. In particular, the probability
of decimating a field, $(1-\alpha\overline{\mu}_{0}){\mathcal{R}}_{\beta}(0)$,
is reduced with increasing dissipation and vanishes for clusters with
finite moment $\mu>1/\alpha$. Thus, a critical fixed point solution
is impossible, and the infinite randomness critical point found in
the dissipationless case \cite{motrunich-ising2d,pich-etal-prl98}
is destroyed by smearing. Moreover, the weakly disordered Griffiths
phase is replaced by the inhomogeneously ordered ferromagnet. Note
that Ohmic dissipation also suppresses the quantum Griffiths singularities
at the percolation quantum phase transition \cite{senthil-sachdev-prl96}
in a \emph{diluted} transverse-field Ising model \cite{hoyos-vojta-prb06}.
However, the percolation transition remains sharp because it is driven
by the critical geometry of the lattice. 

Our results for a dissipative \emph{Ising} magnet must be contrasted
with the behavior of systems with \emph{continuous} O($N$) symmetry.
While large Ising clusters freeze in the presence of Ohmic dissipation,
O($N$) clusters continue to fluctuate with a rate exponentially small
in their moment \cite{vojta-schmalian-prb05}. This leads to a sharp
transition controlled by an infinite-randomness critical point in
the same universality class as the dissipationless random transverse-field
Ising model \cite{hoyos-kotabage-vojta}. All these results are in
agreement with a classification of weakly disordered phase transitions
according to the effective dimensionality of the rare regions \cite{vojta-review06}.
If their dimension is below the lower critical dimension $d_{c}^{-}$
of the problem, the behavior is conventional; if it is right at $d_{c}^{-}$,
the transition is of infinite-randomness type; and if it is above
$d_{c}^{-}$, finite clusters can order independently leading to a
smeared transition. 

To the best of our knowledge, this work is the first quantitative
analytical theory of a smeared phase transition. The results directly
apply to quantum phase transitions in disordered systems with discrete
order parameter symmetry and Ohmic damping. Our renormalization group
approach should be broadly applicable to a variety of disordered dissipative
quantum systems such as arrays of resistively shunted Josephson junctions
\cite{chakravarty-etal-prl86,mpafisher-prl86}. 

This work was supported by the NSF under Grants Nos. DMR-0339147 and
DMR-0506953, by Research Corporation, and by the University of Missouri
Research Board. Parts of the research have been performed at the Aspen
Center for Physics. 

\vspace*{-3mm}

\bibliographystyle{apsrev}
\bibliography{/home/hoyos/Documents/referencias/referencias}

\begin{thebibliography}{24}
\bibitem{foot1} The precise value of $p$ is unimportant, it will not appear in the RG recursion relations.
\bibitem{foot2} There are other, more singular, fixed points but they cannot be reached from nonsingular initial distributions.
\end{thebibliography}

\end{document}